# Splashing phenomena of room temperature liquid metal droplet striking on the pool of the same liquid under ambient air environment

Haiyan Li[1], Shengfu Mei[1], Lei Wang[1], Yunxia Gao[1], and Jing Liu[1,2]*

[1] Beijing Key Lab of CryoBiomedical Engineering & Key Laboratory of Cryogenics,

Technical Institute of Physics and Chemistry,

Chinese Academy of Sciences, Beijing 100190, P. R. China

[2] Department of Biomedical Engineering, School of Medicine,

Tsinghua University, Beijing 100084, P. R. China

*Address for correspondence:

Dr. Jing Liu

Beijing Key Lab of CryoBiomedical Engineering

   & Key Lab of Cryogenics,

Technical Institute of Physics and Chemistry,

Chinese Academy of Sciences,

Beijing 100190, P. R. China

E-mail: jliu@mail.ipc.ac.cn

Tel. +86-10-82543765

Fax: +86-10-82543767

**Abstract:** In this article, the fluid dynamics of room temperature liquid metal (RTLM) droplet impacting onto a pool of the same liquid in ambient air was investigated. A series of experiments were conducted in order to disclose the influence of the oxidation effect on the impact dynamics. The droplet shape and impact phenomenology were recorded with the aid of a high-speed digital camera. The impact energy stored in the splash structures was estimated via a theoretical model and several morphological parameters obtained from instantaneous images of the splash. It was observed that the droplet shape and the splashing morphology of RTLM were drastically different from those of water, so was the impact dynamics between room temperature LM pool and high temperature LM pool. The energy analysis disclosed that the height of the jet is highly sensitive to the viscosity of the fluid, which is subjected to the oxidation effect and temperature effect simultaneously, and thus perfectly explained the phenomena. These basic findings are important for the application of RTLM in a series of newly emerging technologies such as liquid metal based spray cooling, ink-jet printed electronics, interface material painting and coating, metallurgy, and 3D packages, etc.

## 1. Introduction

The room temperature liquid metal (RTLM) or alloy generally has high thermal and electrical conductivity, as well as small vapor pressure. Some metal fluids even could stay at liquid phase within an extensive temperature range, say from their melting points near or far below room temperature until upwardly to above 2000 ℃ (Stein 1966). With such particular characteristics distinguishing from conventional liquids, the RTLMs were recently found highly desirable in a series of emerging important areas. For example, the gallium-based thermal interface material (Gao and Liu 2012) and RTLMs printed electronics (Gao et al. 2012; Li et al. 2012; Zheng et al. 2013) revealed their unconventional merits when a small amount of oxides were included. Among the many properties, the outcome of a RTLM droplet with oxides impacting onto a surface (dry, wet, liquid pools, etc.) is especially a fundamental fluid dynamic problem, which is not only of scientific interest but also significant for quite a few newly emerging practices, such as ink-jet printed electronics, spray cooling, interface material painting and coating, enhancing boiling, metallurgy, and 3D packages (Antkowiak et al. 2011).

Impact of droplets have been an important topic for many years, and some typical trends have been identified for water and aqueous solutions in preceding papers (Manzello and Yang 2002; Sabadini and Alkschbirs 2004; Yarin 2006; Pan and Hung 2010; Lagubeau et al. 2012). However, up to now, only very few studies were ever investigated on the impact dynamics of RTLMs. Hsiao et al. (Hsiao et al. 1988) reported the onset of central jet using the experimental data for mercury. As is known to all, mercury is subject to the inherent toxicity, which hinders its large-scale applications in the above fields. On the other hand, unlike most RTLMs, mercury is a kind of special material that is not apt to form oxide skin when exposed to the atmosphere. In such sense, the outcome of mercury would not be taken as a representative of RTLMs. In fact, among the few works regarding the metal droplet impact dynamics, oxidation effect is usually avoided by protective atmosphere for simplicity (Pasandideh-Fard et al. 1998). This however did not reflect the practical situation since most applications are in ambient air. Due to the lack of systematic information within the open literature on the impact morphology of the RTLMs with oxides, which are expected to have subtly different physics than that of water and aqueous solutions, it is very necessary to carry out a fundamental investigation on this important issue. The motivation for the present study is to explore the new phenomena that may arise when the RTLM droplet impacting on the liquid surface in ambient air, and thus disclose the influence of oxidation effect on the impact dynamics. The measurement is conducted for the normal impacts of a single RTLM droplet striking onto a pool of the same liquid to provide the basis for more complex situations, e.g. multiple droplets, oblique impact, and so on.

**2. Experimental methods**

2.1. Sample preparation

Gallium-indium alloy ($GaIn_{24.5}$) is selected as the experimental fluids, since it has a melting temperature of 15.5 ℃, which is far below the room temperature of 25 ℃. Its density, surface tension and dynamic viscosity are 6280 kg/m$^3$, 0.624 N/m and $1.6956 \times 10^{-3}$ Pa s, respectively, at temperature of 25 ℃ (Morley et al. 2008), which are larger than those of water. $GaIn_{24.5}$ was prepared from gallium and indium metals with purity of 99.99 percent. Such raw materials with a weight ratio of 75.5:24.5 were added into the beaker and heated at 100 ℃. Then a magnetic stirrer was utilized to stir the mixture after they were all melted to achieve uniform mixing. Considering the fact that $GaIn_{24.5}$ is easily subjected to oxidation, the cleaning processing prior to the experiment is necessary to assure the definite initial state. For this purpose, 10 ml 30 % NaOH solution was added slowly to the alloy, and the mixture was stirred at room temperature for short periods of time until the contents of the beaker comprised obvious two-phase structure, i.e., an aqueous phase and a metallic phase. Then, the $GaIn_{24.5}$ alloy was extracted from the mixture by a syringe.

2.2. Experimental apparatus

The experiments were performed using a setup consisting of the droplet generator, target surface, heating element, and imaging system, as shown in Fig. 1. The $GaIn_{24.5}$ liquid pool was contained in a transparent plastic cylinder, with 85 mm in diameter and 15 mm in height. The syringe (10 ml) filled with cleaned $GaIn_{24.5}$ was mounted horizontally in a syringe pump (Longer LSP10-1B). A flat tipped stainless steel needle attached to a travelling vernier (0.01 mm accuracy) was positioned vertically above the center of the liquid container, with its termination connected to the outlet of the syringe through silicon rubber tubing. Thus, the droplet was generated using the syringe pump programmed to dispense the liquid at a rate of 0.5 ml/min, and formed at the tip of the needle, then detached off under its own weight. Droplet impact dynamics were captured using a high-speed digital camera (IDT, NR4-S3, USA) at 5000 frames per second with a resolution of 512 × 512 pixels and a span of 2.825 s. The camera was fitted with a Nikon 85-mm micro lens to obtain the required spatial resolution to capture the impact, and it was aligned at an angle $\theta = 10°$ with respect to the horizontal to acquire the status of height and horizontal surface simultaneously. An exposure time of 198 μs was synchronized with a 1000 W tungsten light, and a high-power LED lamp was taken as compensatory light to adequately illuminate the impact region. The light sources were kept as far away from the test surface as possible to mitigate heating of the droplet and liquid surface. The lights were switched on only in the process which began from the dropping of the droplet and ended after droplet impact, and the total time when the lights kept on was no more than 5 s per experiment. The camera was connected to a PC, which contained the corresponding control software. The captured images were recorded using the software, and then the impact velocity and geometric parameters of the droplet and splashing can be quantitatively extracted from the instantaneous images with the aid of the software. In view of the 10° down angle of the camera, parameters in the vertical direction including impact velocity and splashing height etc. measured from the images were modified through dividing by $\cos 10°$. It should be noted that Fig. 1 (a) was used to clarify the effects of droplet size and the impact velocity at room temperature on the impact dynamic. Besides that, a heating unit consisting of two heating rods and an aluminum block was added to test the influence of pool temperature, as shown in Fig. 1 (b). The heating rods were powered from a voltage regulator. Considering the poor heat resistance of the plastic container, a stainless steel one was employed instead in this case. The temperatures for the pool and block were measured simultaneously using Type T thermocouples.

## 3. Results and discussion

### 3.1. Droplet shapes

As is noticed, most of the previous literatures were based on the assumption that the droplets were spherical, although the shape of droplets moving through a fluid will always be rendered slightly ellipsoidal by aerodynamic forces (Rein 1993). However, this is not the case in our experimental work. As is well known, common RTLMs like gallium and its alloys tend to form oxide skin in the room temperature atmosphere, which will prevent further oxidation. It has been disclosed that the surface of gallium-indium alloy droplet became oxidized during the dispensing action in less than 0.25 s (Liu et al. 2012), and the component of oxide skin was mainly gallium oxide (Dickey et al. 2008). Zrnic and Swatik (1969) have ever reported their measurements of the surface tension of a pendant droplet of EGaIn ($GaIn_{25}$) at ambient conditions, which was ~624 mN/m, whereas the surface tension of $GaIn_{25}$ without oxide skin was ~435 mN/m (Dickey et al. 2008). Obviously, the influence of the ambient gas cannot be neglected for RTLMs although this simplification is normally employed for water's case. Fig. 2 (a) shows the effect of ambient gas on RTLM droplet vividly. During the falling, the oxide skin prevents the droplet from releasing the surface energy freely, thus, the tail is held until the impact. In order to further verify the role of oxidation for the shape of the droplet, the experiment is repeated in the NaOH solution as shown in Fig. 2 (b). A small tail is still observed at the initial stage of the detachment, but it disappears gradually with the falling as anticipated. NaOH solution has been previously presented to remove the oxides, and it also provides an inert environment for $GaIn_{24.5}$ (Taylor and Rancourt 1998). For comparison, the free falling of a deionized water droplet in ambient air is illustrated in Fig. 2 (c), attachment is also observed at the initial moment of detachment but the water droplet contracts and forms a spherical shape within a short time due to surface tension. This outcome is similar with that exhibited in Fig. 2 (b), except for the shape of the pendant droplet, which is believed as the result of difference of the buoyancy. All these images are obtained with the same inner diameter of needle of 1 mm, and falling height of 900 mm.

In view of the non-spherical droplet, equivalent droplet diameter is introduced, which is defined as the droplet diameter when taking the droplet as sphere, which can be calculated from

$$\pi d_e^3 / 6 = m / \rho \qquad (1)$$

i.e.

$$d_e = \sqrt[3]{6m/\pi\rho} \qquad (2)$$

where $m$ is the droplet mass, which can be obtained by weighing the pendant droplet. The calculated equivalent droplet diameter $d_e$ and the measured droplet horizontal width $d$ from the images for the same size needle are compared in Fig. 3. Here, the droplet horizontal width $d$ is measured at the maximum horizontal width of the droplet from the image just prior to the impact as observed. The error bars represent a relatively standard uncertainty in determining the $d_e$ and $d$, with the standard deviation of 0.09 mm and 0.04 mm, i.e. maximum uncertainty of 1 % and 3 %, respectively. It can be noted that $d_e$ and $d$ are close to each other when the small size needle is employed. But their discrepancy enlarges with the increase of the needle size.

Normally, in experiments the impact energy is a most important parameter, among which, surface energy is a critical contributing one for droplet shape. Therefore, the dependence on the Weber ($We$) number of the droplet will be discussed in the following. $We$ number is the ratio of the kinetic energy to the surface energy of the impinging droplet, which is defined as

$$We = \rho V^2 D / \sigma \tag{3}$$

where $V$ is the droplet impact velocity, $D$ is the characteristic length, and $\rho$ and $\sigma$ respectively the density and surface tension of the liquid. The density and surface tension in this study are based on properties of the impacting droplet extracted from the previous literature.

On the other hand, in view of the particularity of RTLMs, the oxidation effect of which can improve the viscosity of RTLM droplet (Gao et al. 2012), the viscosity is not included in the $We$ number. Thus, the Reynolds ($Re$) number relating the inertia and the viscosity is introduced here to describe the impact process, which is defined as

$$Re = \rho V D / \mu \tag{4}$$

where $\mu$ is the dynamic viscosity.

During the falling process, gravity is also an important parameter, thus Froude ($Fr$) numbers defined as the ratio of inertia to gravitational forces is of significance, which is expressed as

$$Fr = \frac{V}{\sqrt{gD}} \tag{5}$$

where $g$ is the gravitational acceleration.

Based on Eqs. (3–5), it can be concluded that characteristic length and impact velocity are key parameters for the specific fluid. The size of the droplet is varied through use of different-sized needles, while the impact velocity is adjusted by varying the falling height of the impacting droplet, i.e. the distance between the needle tip and the surface of the liquid pool. Table 1 and Table 2 list main parameters for the cases of varied droplet size and varied impact velocity respectively. In tables, $d_i$ is the inner diameter of needle, $H$ is the released height, $Fr_e$, $We_e$ and $Re_e$ are the parameters taking $d_e$ as characteristic length, while $Fr$, $We$ and $Re$ are the parameters taking $d$ as characteristic length. The uncertainties in the experiments are estimated as ±5%. The calculations indicate that there is no big difference for dimensionless numbers whether $d_e$ or $d$ is utilized. Thus, there is an acceptable deviation when taking the droplet as sphere.

**Table 1** Main experimental parameters ($V = 4.2$ m/s)

| $d_i$ (mm) | $d$ (mm) | $d_e$ (mm) | $Fr$ | $Fr_e$ | $We$ | $We_e$ | $Re$ | $Re_e$ |
|---|---|---|---|---|---|---|---|---|
| 0.4 | 2.7 | 2.7 | 25.81989 | 25.96096 | 479.3331 | 474.138 | 42000 | 41544.8 |
| 1 | 3.4 | 4.1 | 23.00895 | 21.07214 | 603.6046 | 719.6622 | 52888.89 | 63058.05 |
| 1.6 | 3.9 | 4.5 | 21.48345 | 20.01671 | 692.37 | 797.5556 | 60666.67 | 69883.21 |
| 2.2 | 4.3 | 5.2 | 20.45983 | 18.62492 | 763.3823 | 921.2075 | 66888.89 | 80717.81 |

**Table 2** Main experimental parameters ($d = 3.4$ mm, $d_e = 4.1$ mm)

| $H$ (mm) | $V$ (m/s) | $Fr$ | $Fr_e$ | $We$ | $We_e$ | $Re$ | $Re_e$ |
|---|---|---|---|---|---|---|---|
| 300 | 2.4 | 13.14797 | 12.04117 | 197.0954 | 234.994 | 30222.22 | 36033.53 |
| 500 | 2.9 | 15.88713 | 14.54974 | 287.7729 | 343.1076 | 36518.52 | 43540.51 |
| 700 | 3.5 | 19.17412 | 17.56003 | 419.1699 | 499.7703 | 44074.07 | 52548.9 |
| 900 | 4.2 | 23.00895 | 21.07204 | 603.6046 | 719.6693 | 52888.89 | 63058.68 |

3.2. Impact process

Repeated experiments were performed at various droplet sizes, impact velocities and pool temperatures. Since experiments at various droplet sizes and impact velocities conducted at room temperature display similar qualitative trends, Fig. 4 is employed as a representative of the two cases. In this series, the equivalent droplet diameter and impact velocity are 4.1 mm and 4.2 m/s respectively at room temperature for an $Fr_e$ number of 21, $We_e$ number of 720 and $Re_e$ number of 63058. The image just prior to the observed impact was chosen as the onset time image in each sequence. In the first milliseconds after the droplet impact, the inertia drives the fluid to form an expanding cavity and the bulk fluid around the impact point flows upward because of continuity. Then very small droplets are ejected from the top of the cavity at 2.6 ms, which is termed as the characteristic of crown. According to Engel's work, the crown is mainly composed of target liquid, but usually also contains some of the droplet liquid that lines the cavity (Engel 1966). In addition, it is noted that the ejected droplets are not spherical, but fusiform, which is believed to be caused by oxidation of the droplet liquid and surface target liquid. The crown continues to grow in height and size with time producing more secondary droplets, and reaches its maximum of 9 mm, measured from the pool surface to the upper edge of the cavity, after about 20 ms, and then starts to subside until changes into a wave swell at 50 ms. This leads to the formation of the central jet that carries droplet liquid at its top and reaches a maximum of 28 mm at about 110 ms. With the jet rising the neck becomes narrow due to the inward pull of the surface tension, but no droplet separates from the tip of the jet. Then, the jet sinks to form a second crater at about 200 ms after the initial droplet impact. As the droplet continues its downward motion a smooth capillary wave front moves outwards. Splashing heights for different droplet sizes and impact velocities are compared in Fig. 5 (a) and (b) respectively, and the interval between every two data is 10 ms. All the measurements are taken relative to the initial pool surface level. It is observed that the maximum crown (minor peak) and jet (major peak) height basically increase with the increased droplet size and impact velocity, and increases of droplet size and impact velocity also lead to the extension of the time from the impact to the formation of a second crater.

As for the cases under various pool temperatures, the equivalent droplet diameter and impact velocity were still fixed at 4.1 mm and 4.2 m/s, while the liquid pool was heated from 50 ℃ to 200 ℃. Fig. 6 shows the impact dynamics at 200 ℃. At a time of 20 ms after impact, the crown reaches the maximum of 8 mm. After the crown collapses at 50 ms, a slender high-speed thin jet is ejected upward, whose height reaches a maximum value of 40 mm at about 85 ms. Here, the jet height is defined as the height the liquid jet rises above the free surface prior to jet breakup. It is clearly larger than the maximum height reached by the jet of the case as shown in Fig. 4. Interestingly, two necks are formed gradually, partitioning the jet into three parts. Then the upper two parts pinch off together at about 90 ms. This should be caused by the decreasing drag that prevents the liquid from taking apart, and is also apprehensible from the energy aspect of viscous dissipation. The detached tip droplet rises further while the jet quietly subsides into the bulk liquid. The droplet detached later begins leaving the field of view (100 ms). Except for these features the whole process proceeds in the usual way. In these cases, the formation of a crown and/or the rise of a central jet are usually regarded as the main features of splashing, and secondary droplets are produced from the rim of crown and the tip of central jet since the crown and central jet are normally unstable (Okawa et al. 2006). A quantitative comparison for different pool temperatures is undertaken by comparing the splashing height as a function of time in Fig. 7. Note that the jet height does not include tip droplets that break off from the jet. Thus when the droplet breaks off from the jet, the jet height as a function of time appears to jump discontinuously. This observation was also made by Morton et al. in their experiments for water (Morton et al. 2000). Furthermore, it does not seems that the variation of spashing height and the moment of breaking off obey a certain law with the increase of pool temperature, but the maxmum height is definitely higher than the case without heating. It may be explained that heating contributes to the impact energy.

The energy stored at different stages of the splash can be roughly estimated using a simple model, on the basis of energy conservation and a simplification of the impact structures geometry. In a first approach, the evolution of the splash can be analyzed by considering two stages (Fig. 8): (a) The moment of impact; (b) The maximum amplitude of the jet. For simplicity, the droplet can be considered to be a sphere, and the jet a liquid cylinder.

The energy conservation of the two stages can be expressed as follows

$$Q_{im} = Q_j + Q_D + Q^* \tag{6}$$

where $Q_{im}$ is the droplet impact energy, $Q_j$ is the energy required to produce the jet, $Q_D$ is the dissipated viscous energy and $Q^*$ represents other dissipative energies (as heat and sound).

The droplet impact energy ($Q_{im}$), can be determined from the kinetic and potential energies (gravitational and surface), as expressed in Eq. (7)

$$Q_{im} = mg(H-h) + mg\frac{d_e}{2} + 4\pi\left(\frac{d_e}{2}\right)^2 \sigma \tag{7}$$

where $h$ is the depth of the target liquid.

The energy necessary to create the jet ($Q_j$) can be determined using Eq. (8)

$$Q_j = \frac{\pi}{2}\rho g\left(\frac{d_j}{2}\right)^2 H_j^2 + \pi\left(\frac{d_j}{2}\right)\left[\left(\frac{d_j}{2}\right) + 2H_j\right]\sigma \tag{8}$$

where the first and second terms in the Eq. (8) are the gravitational and surface energies, respectively; $d_j$ and $H_j$ are the diameter and heights of the jet, which were determined from "instantaneous pictures" of the splash in stage two. The disconnected droplets were not considered during determination of $H_j$.

Given the conditions that the equivalent droplet diameter and impact velocity is 4.1 mm and 4.2 m/s at room temperature, the parameters, calculated energies for the impact droplet and jet are shown in Table 3. The calculated impact energy is $Q_{im} = 1.9 \times 10^{-3}\,\text{J}$, and the energy required to produce the jet is $Q_j = 9.6 \times 10^{-4}\,\text{J}$. Approximately 50 % of the impact energy is stored in the jet. In other words, nearly half of the impact energy is released due to viscous dissipation as other dissipative energies only account for a small fraction. Considering the dissipated viscous energy mainly depends on the viscosity (Sabadini and Alkschbirs 2004), the viscosity may thus be the major cause leading to the difference between the room temperature and high temperature cases. For RTLMs, the viscosity is subjected to the oxidation and temperature effects simultaneously. The combined effect between them leads to the particular impact phenomenology.

**Table 3** Parameters and energies obtained from the splash in GaIn$_{24.5}$

| $m$ | $H$ | $h$ | $\sigma$ | $d_e$ | $Q_{im}$ | $\rho$ | $d_j$ | $h_j$ | $Q_j$ |
|---|---|---|---|---|---|---|---|---|---|
| (g) | (mm) | (mm) | (N/m) | (mm) | ($10^{-3}$J) | (kg/m$^3$) | (mm) | (mm) | ($10^{-3}$J) |
| 0.22 | 900 | 15 | 0.624 | 4.1 | 1.9 | 6280 | 5.8 | 27.5 | 0.96 |

3.3. Splashing shapes

Fig. 9 shows the comparison of the representative droplets and splashing shapes over the process for both GaIn$_{24.5}$ and deionized water with the same inner diameter of needle of 1 mm, and falling height of 900 mm. In water's case (Fig. 9 (a)), spherical droplet and secondary droplet are observed no surprisingly, which is consistent with Manzello' work (Manzello and Yang 2002). However, the dynamic viscosity (1.7×10$^{-3}$ m$^2$/s) of GaIn$_{24.5}$ is higher than that of water (1.002 ×10$^{-3}$ Pa s), and even much higher when subjected to surface oxidation, thus secondary droplet is hard to form under the effect of elevated viscous force, as shown in Fig. 9 (b). With the increase of the temperature of the pool, as illustrated in Fig. 9 (c), the viscosity of GaIn$_{24.5}$ droplet decreases after coalescence, and the viscous force draging the liquid weakens, thus the secondary droplet forms. Yet, the most dramatic difference from the water's case is that the tip detached is not spherical, but fusiform. This is believed to be caused by the serious surface oxidation of GaIn$_{24.5}$ at high temperature. The images are obtained with the same inner diameter of needle of 1 mm, and falling height of 900 mm.

The influences of the falling height, inner diameter of needle and pool temperature on the GaIn$_{24.5}$ droplet shapes are depicted in Fig. 10. It can be found that the tail is apparently more pronounced as inner diameter of needle increases, while the influences of falling height and pool temperature appear much less. The former is in accordance with the fact that larger inner diameter of needle leads to the larger surface area of droplet, thereby more oxides formation around the surface. The raised falling height prolongs the falling time of the droplet, but the oxide skin prevents the droplet from further oxidation. Thus the content of oxides does not increase with the time exposing to the ambient air; while the pool temperature is independent of the temperature of metal droplet during the early stages of droplet impact, i.e. the temperature of metal droplet equals to its initial temperature (Aziz and Chandra 2000).

Furthermore, as is well known, when a cold water droplet is brought into direct contact with a hot water pool, it is possible that the cooler water droplet may vaporize so rapidly that an explosion may occur. These explosions have been termed as vapor explosions, explosive boiling, or rapid vapor explosions (Manzello et al. 2003). In contrast, it is not a concern for the RTLMs with boiling point of above 2000 ℃. This is also regarded as the superiority of RTLMs over water for high temperature applications.

## 4. Conclusions

In this study, the splashing dynamics of RTLM droplets with oxide skin impinging vertically on a pool composed of the same liquid were visualized and quantitatively investigated using a high speed camera. The influences of the droplet size, impact velocity and pool temperature on the fluid striking dynamics were revealed on the basis of experimental observations, image data processing, and mechanism interpretation. It was found that the droplet size and the impact velocity displayed similar proportional trends with respect to the splashing height, but did not accompany with the secondary droplet separation; while the increase of pool temperature dramatically intensified the splashing effect, with the fusiform secondary droplet detached from a central jet. The $Fr_e$ number covered in this work is from 20 to 26, while $We_e$ number is from 479 to 763, and $Re_e$ number from 42000 to 66889, which are all calculated with the equivalent droplet diameter introduced specially for the non-spherical droplet. It has been disclosed by the present work that there exist fundamentally distinct differences between the impact behavior of $GaIn_{24.5}$ droplets and that of deionized water droplets. The reason can be attributed to the oxide skin of the RTLM which would significantly affect the appearance of the droplets and the splashing morphology. The energy analysis further confirmed the speculation. These findings are expected to be important for the application of RTLM in a series of newly emerging technological fields such as liquid metal based ink-jet printed electronics, spray cooling, interface material painting and coating, enhancing boiling, metallurgy, 3D packages, and so on.

**Acknowledgments**

The authors thank Prof. Yixin Zhou for valuable discussions.

**Figures and captions:**

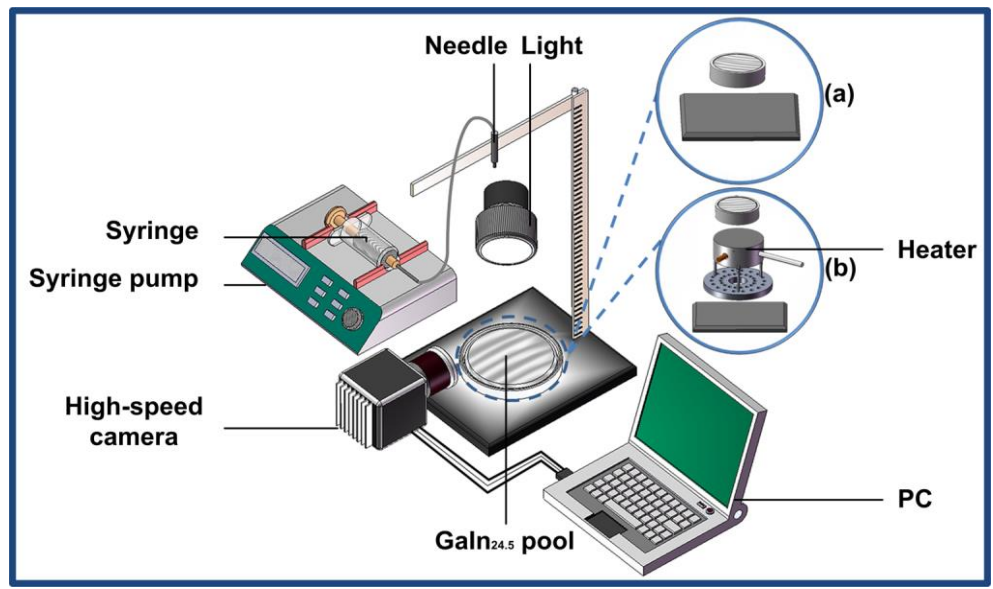

**Fig. 1** Schematic of the experimental setup: **a** Without heating unit; **b** With heating unit under the pool.

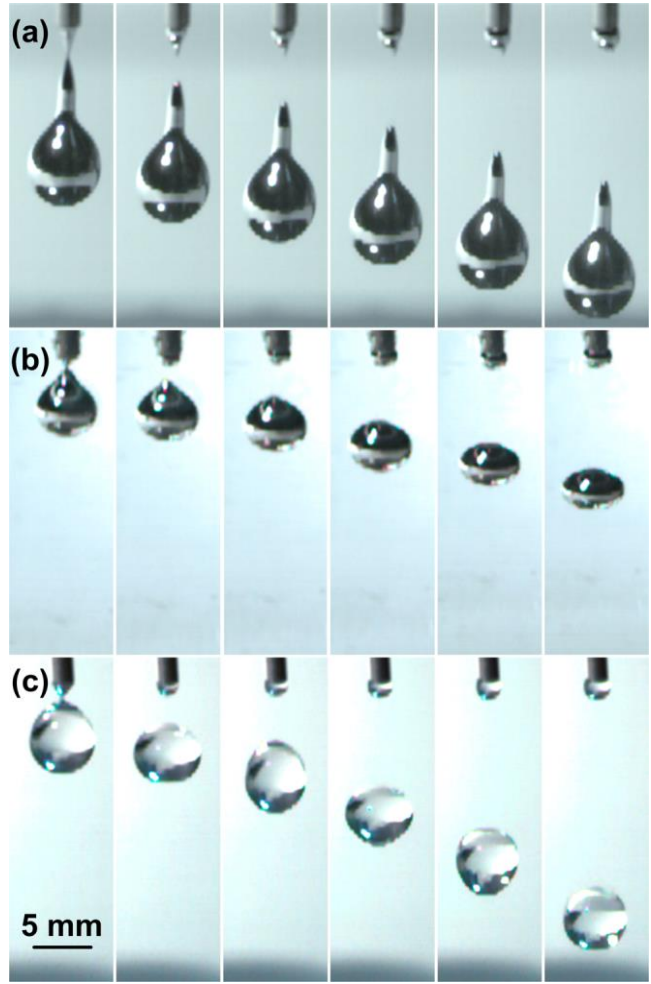

**Fig. 2** Images of free falling: **a** GaIn$_{24.5}$ droplet in ambient air; **b** In the NaOH solution; **c** A deionized water droplet in ambient air.

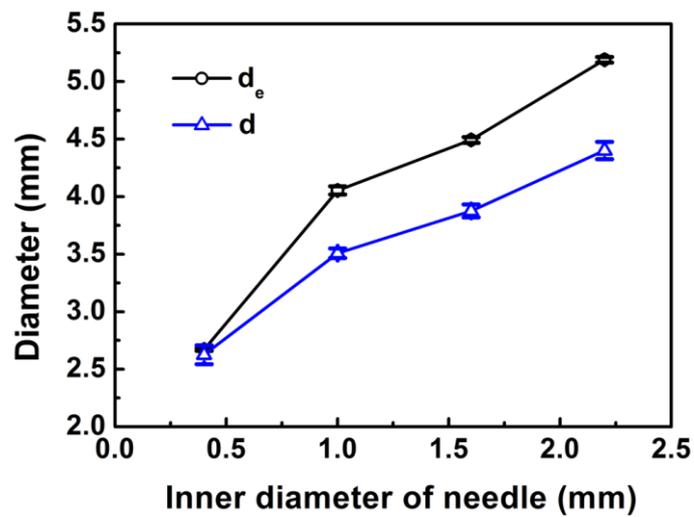

**Fig. 3** Comparison of the equivalent droplet diameter $d_e$ and the droplet horizontal width $d$ measured directly from the images for the same size needle.

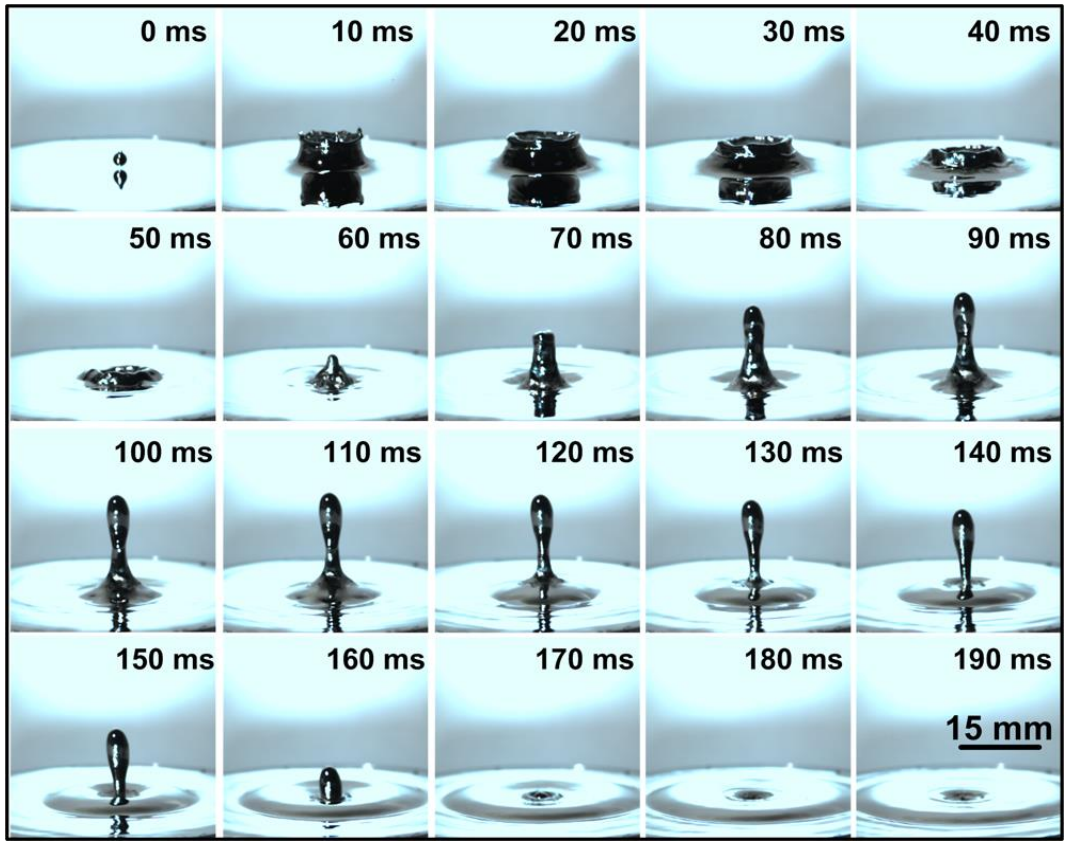

**Fig. 4** Instantaneous images of impact process, obtained with a GaIn$_{24.5}$ droplet of the horizontal width $d_e = 4.1$mm impacting a pool of the same liquid at impact velocity $V = 4.2$m/s and room temperature.

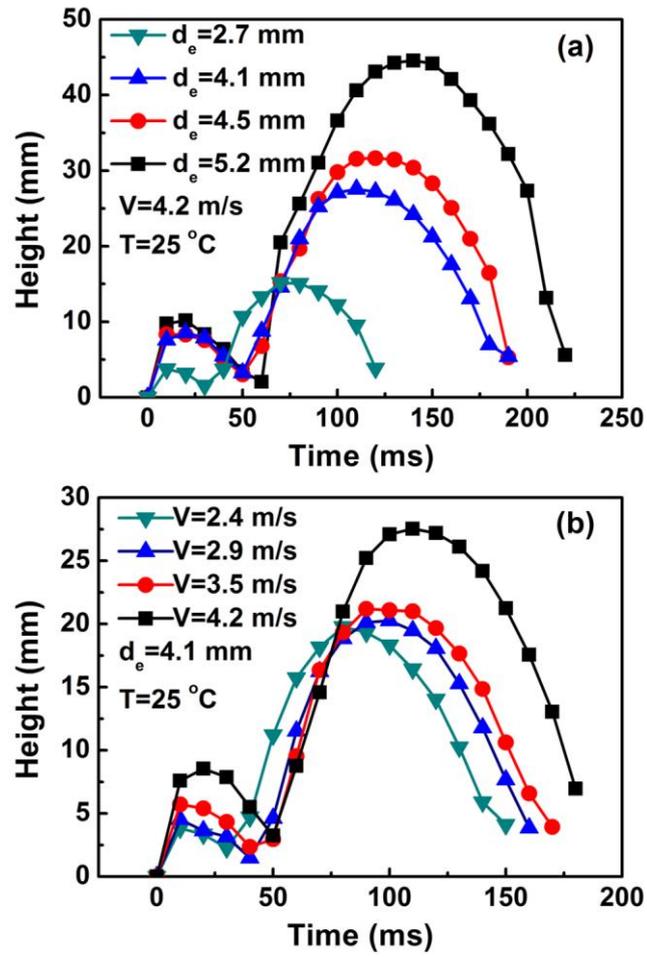

**Fig. 5 a** Comparison of the splashing heights for different droplet horizontal width; **b** Comparison of the splashing heights for different impact velocities.

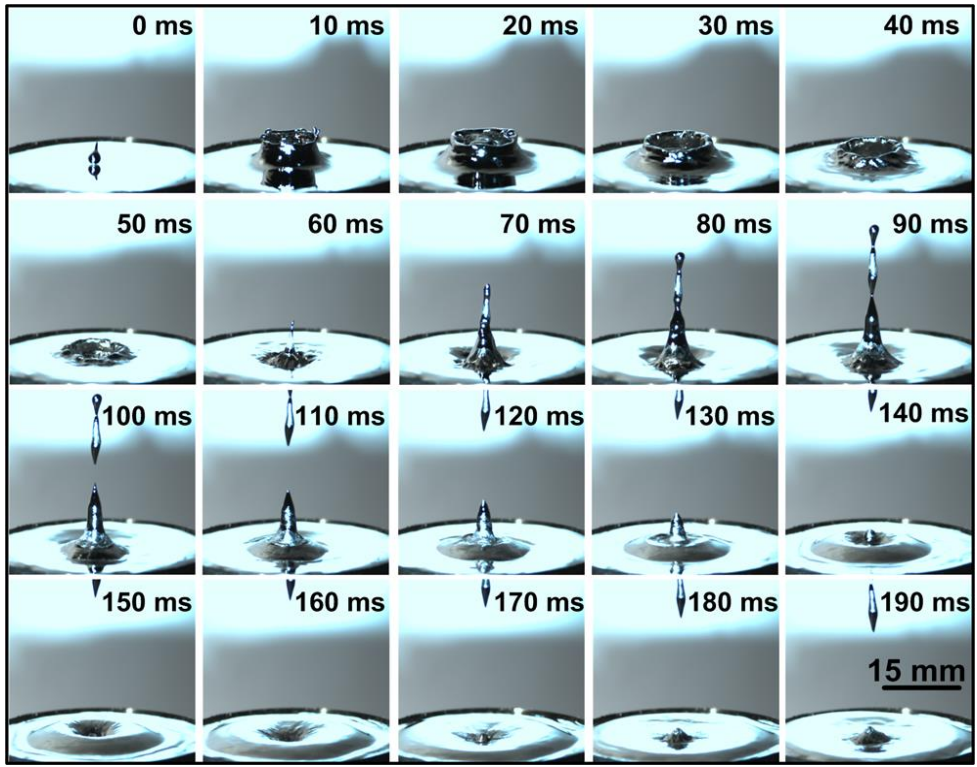

**Fig. 6** Instantaneous images of impact process, obtained with a GaIn$_{24.5}$ droplet of the horizontal width $d_e = 4.1$mm impacting a pool of the same liquid at velocity $V = 4.2$m/s and pool temperature $T = 200°C$.

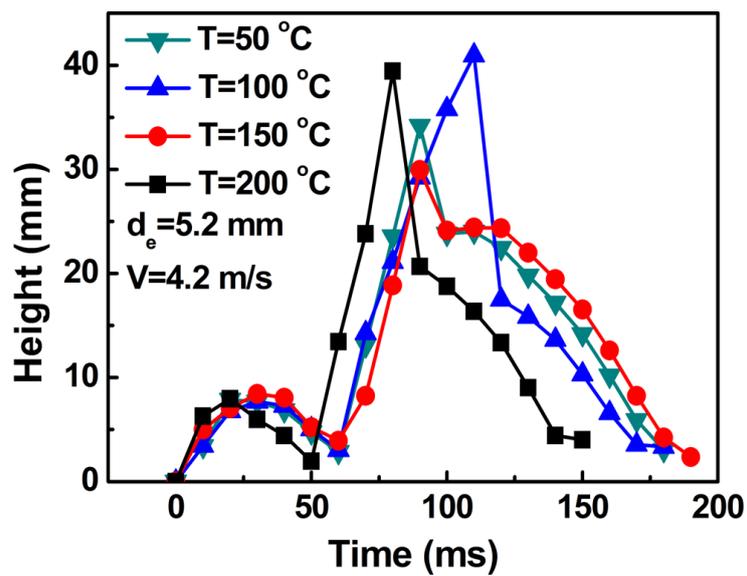

**Fig. 7** Comparison of the splashing heights for different pool temperatures.

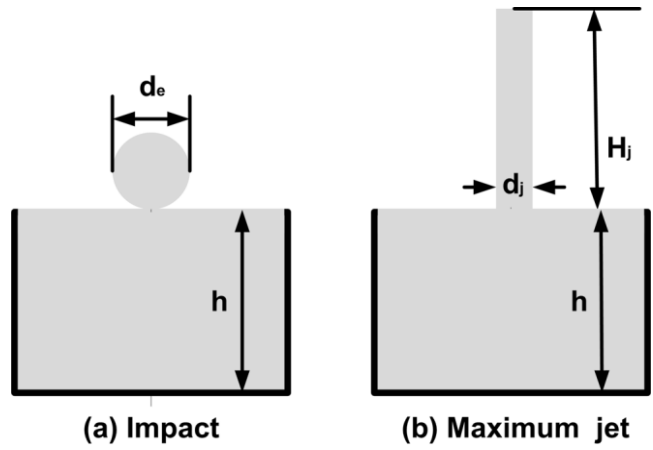

**Fig. 8** Schematic representation of two stages of the splash: **a** The moment of the impact; **b** The maximum height of the jet.

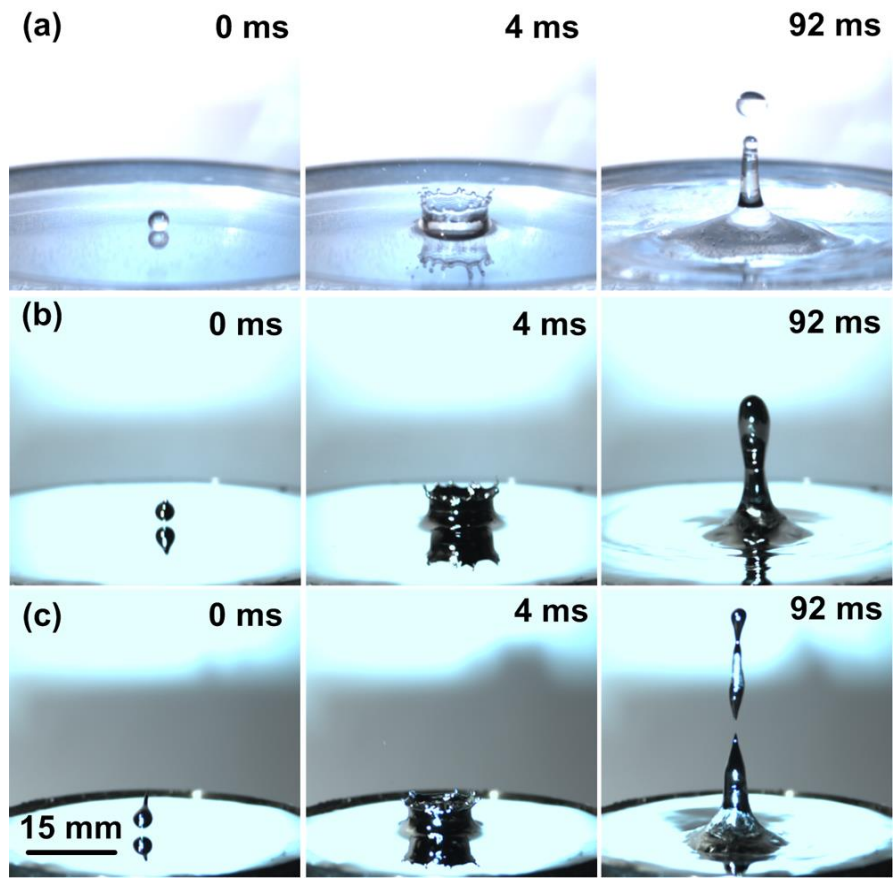

**Fig. 9** Comparison of droplet and splashing shapes during the splashing process at the same moment: **a** Deionized water (25 ℃); **b** GaIn$_{24.5}$ (25 ℃); **c** GaIn$_{24.5}$ (200 ℃).

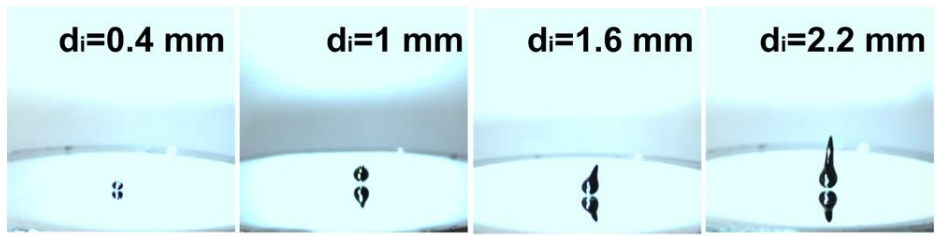
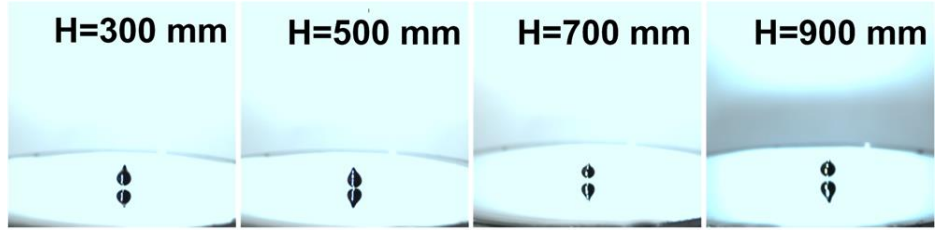
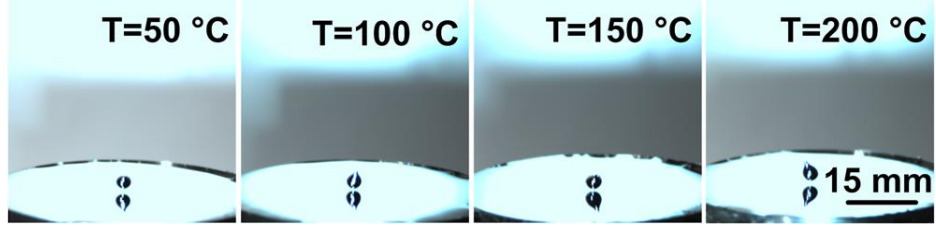

**Fig. 10** Comparison of GaIn$_{24.5}$ droplet shapes at the initial moment ($t=0$ms) for different inner diameter of needle, falling height and pool temperature.